\documentclass[prb,twocolumn,showpacs,preprintnumbers,amsmath,amssymb]{revtex4}

\usepackage{graphicx}
\usepackage{dcolumn}
\usepackage{bm}

\def\R{\partial}
\def\th{\theta}

\def\ep{\varepsilon}

\def\Ps{\varPsi}
\def\Ps{\Psi}

\begin{document}
\title{Interacting quantum rotors in oxygen-doped germanium}
\author{Hiroyuki Shima and Tsuneyoshi Nakayama}
\affiliation{Department of Applied Physics, Graduate School of Engineering,
Hokkaido University, Sapporo 060-8628, Japan}
\date{\today}

%%%%%%%%%%%%%%%%%%%%%%%%%%%%%%%%%%%%%%%%%%%%
%
%  Abstract
%
%%%%%%%%%%%%%%%%%%%%%%%%%%%%%%%%%%%%%%%%%%%%

\begin{abstract}

We investigate the interaction effect between oxygen impurities 
in crystalline germanium on the basis of a quantum rotor model.
The dipolar interaction of nearby oxygen impurities engenders non-trivial low-lying excitations,
giving rise to anomalous behaviors for oxygen-doped germanium (Ge:O) below a few degrees Kelvin.
In particular, it is theoretically predicted
that Ge:O samples with oxygen-concentration of 10$^{17-18}$cm$^{-3}$ show
(i) power-law specific heats below 0.1 K,
and (ii) a peculiar hump in dielectric susceptibilities around 1 K.
We present an interpretation for the power-law specific heats, 
which is based on the picture of local double-well potentials
randomly distributed in Ge:O samples.
\end{abstract} 

%\keywords{ }

\maketitle

%%%%%%%%%%%%%%%%%%%%%%%%%%%%%%%%%%%%%%%%%%%%
%
\section{Introduction}
%
%%%%%%%%%%%%%%%%%%%%%%%%%%%%%%%%%%%%%%%%%%%%

Rotational states of impurity atoms in crystalline solids
affect the essential properties of solids at low temperatures.
This is because the rotational degree of freedom of an impurity
yields low-energy excitations that manifest themselves such low temperatures.
A typical example is an interstitial oxygen impurity
in covalent semiconductors such as 
Si \;\cite{Bosomworth, Kaneta, ShimuraSi,Kaneta2}, 
Ge \;\cite{Kaiser, Corbett, Clauws},
and GaAs \;\cite{Pesola, Alt, Schneider}.
The microscopic structure around
oxygen impurities in those systems has been thoroughly studied in past.
As-grown Ge, for instance, is usually contaminated by 
more than $10^{17}$ oxygen atoms per cubic centimeter with a spatially random distribution.
Oxygen breaks a covalent bond between two Ge atoms,
establishing a puckered Ge-O-Ge segment.
At low temperatures,
those interstitial oxygen atoms are quantum-mechanically delocalized
in the annulus around the original bond center of Ge-Ge
\;\cite{Gienger,Mayur,Artacho97}.
The rotational degree of freedom of oxygen
yields low-lying excitations in the far-infrared spectral region,
which has been observed in phonon spectroscopy measurements \cite{Gienger}.
The energy spectra deduced from these measurements can be accurately reproduced by
a quantum rotor model \cite{Artacho97,Pajot2000,Coutinho};
Each rotating Ge$_2$O unit within this model is mapped onto a quantum rigid rotor
rotating in a two-dimensional plane that is perpendicular to the Ge-Ge axis.

Since the electronegativity of oxygen is larger than that of germanium,
puckered segments of Ge-O-Ge carry electric dipole moments.
In the dilute limit, 
the interaction between the dipole moments is negligible
because the dipolar interaction decreases with the third power of distance.
However, for actual concentrations of oxygen $\rho \sim 10^{17}-10^{18}$ cm$^{-3}$,
a random distribution of oxygen impurities causes the occurrence of clusters
of nearby dipole moments with a large coupling energy.
If the coupling is strong, such that it is sufficient to interfere with the nearly free rotation of 
individual Ge$_2$O units, the quantum nature of the Ge$_2$O units that are involved in the clusters
is completely different from those of the isolated Ge$_2$O units.
Recently, anomalous dielectric responses in coupled dipolar rotors
have been reported \cite{PRB,PRA},
wherein the anomalies originate from peculiar low-lying excitations of the interacting dipolar rotors.
Therefore, it is expected that
ensembles of clustered Ge$_2$O units 
play an important role in the low-temperature properties of Ge:O;
specific heats and dielectric susceptibilities are cases in point.

The present article theoretically investigates the effect of the dipolar interaction
of rotating Ge$_2$O units on the low-temperature properties of Ge:O.
It is shown that 
the strong dipolar interaction between clustered Ge$_2$O units
results in peculiar low-energy states,
which in turn give rise to a power-law temperature dependence in specific heats $C(T)\propto T^{0.5}$ 
below 0.1 K under our numerical conditions.
Further, dipolar transitions between the low-lying levels of the interacting Ge$_2$O units
engender  a non-trivial maxima in the dielectric susceptibilities $\chi(T)$ around 1 K.
The power-law behavior of specific heats is described by a scenario based on the presence of local double-well potentials
in crystalline Ge.
It should be noted that this picture of {\it crystalline} solids is exceedingly analogous to 
the theoretical model, referred to as the two-level tunneling model, 
which describes the power-law specific heats in {\it amorphous} solids \cite{Esqui}.

The outline of this article is as follows.
In Sec. II, we introduce a quantum rotor model that accounts for 
the quantum rotation of the Ge$_2$O unit
and formulate the Hamiltonian for interacting quantum rotors 
that are coupled via dipolar interaction.
Section III focuses on the low-energy excitations of paired and clustered rotors;
the distribution function $P(\ep)$ of the lowest excitation energy $\ep$
is evaluated in line with the theory of the two-level tunneling systems.
Section IV shows the calculated results of the specific heat $C(T)$ for Ge:O.
The power-law behavior of $C(T)$ is understood
by taking into account the distribution function $P(\ep)$ of strongly coupled rotors.
Section V describes the results of the dielectric susceptibilities $\chi(T)$ of Ge:O.
An anomalous hump in $\chi(T)$ around 1 K is understood by considering 
the selection rule for dipolar transition.
Section VI comprises remarks on the relevance of our findings to other systems.
The final section concludes the present article.

%%%%%%%%%%%%%%%%%%%%%%%%%%%%%%%%%%%%%%%%%%%%
%
\section{System}
%
%%%%%%%%%%%%%%%%%%%%%%%%%%%%%%%%%%%%%%%%%%%%

\subsection{Isolated quantum-rotor model}

%----------------------------------------------------------------------------------
\begin{figure}[bbb]
\vspace{0cm}
\hspace{0cm}
\includegraphics[width=8.5cm]{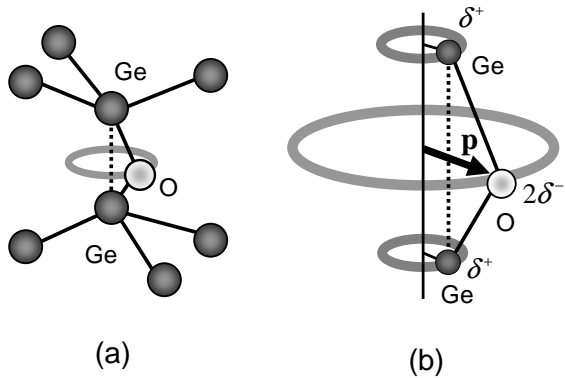}
\vspace{0cm}
\caption{(a) The atomic coordinate of a puckered Ge$_2$O unit
together with six Ge atoms next nearest to the interstitial oxygen.
(b) Microscopic details of a rotating Ge$_2$O unit.
The whole Ge$_2$O unit rotates in phase around an external axis.
The difference of the electronegativity between Ge ($\delta^+$) and O ($2\delta^-$)
causes the occurrence of an electric dipole moment $\bm{p}$
in the direction perpendicular to the rotational axis.}
\label{fig_01}
\end{figure}
%---------------------------------------------------------------------------------

We first describe the microscopic structure of an isolated Ge$_2$O unit,
taking no account of interaction between them.
Figure \ref{fig_01} (a) illustrates the atomic disposition of a puckered Ge-O-Ge segment
together with six Ge atoms next nearest to the interstitial oxygen.
At sufficiently low temperatures, the oxygen atom is quantum-mechanically delocalized around the Ge-Ge axis.
The delocalized oxygen is subjected to repulsive force
from the six next-nearest neighboring Ge atoms,
thus experiencing an azimuthal potential having six-fold rotational symmetry
\cite{Gienger}.
Hereafter, we call it a "hindering potential".

Recent isotope-shift measurements \cite{Artacho97}
have revealed that 
the two Ge atoms neighboring O are not fixed, but loosely bound around an external axis.
As a consequence, 
all three atoms (Ge and O) rotate in phase around the external axis
as shown in Fig.~\ref{fig_01} (b).
The difference of the electronegativity between 
Ge and O causes the occurrence of an electric dipole moment
in the direction perpendicular to the Ge-Ge axis.
This allows us to map a rotating Ge$_2$O unit onto a quantum rotor
having the moment of inertia \cite{Gienger,Artacho97} $I=10.6$ amu$\cdot$\AA$^2$
and an electric dipole moment \cite{foot} $\bm{p}=1$ Debye .

The Hamiltonian $H$ for an isolated rotor consists of two parts:
\begin{equation}
H = K + V.
\label{eq_02}
\end{equation}
The first term $K$ accounts for the kinetic energy of 
the rotor associated with its rotation, given by
\begin{equation}
K = -\frac{\hbar^2}{2I} \frac{\R^2}{\R\th^2}.
\label{eq_03}
\end{equation}
The quantity $E_K=\hbar^2/(2I)$ determines the energy scale of the quantum rotation.
The second $V$ corresponds to a hindering potential,
which reads
\begin{equation}
V = V_0 \cos(6\th+\phi)
\label{eq_04}
\end{equation}
with a constant $\phi$.
In the following, we fix $E_K=2.3$ K and $V_0 = 1.5 E_K$
in accordance with the values deduced from the spectroscopy measurement \cite{Gienger}.
The Schr\"odinger equation for the Hamiltonian (\ref{eq_02})
can be solved analytically
by mapping it onto the Mathieu equation;
the details of these calculations are given in Refs.~\cite{Whittaker} and \cite{Gloden}.

%---------------------------------------------------
\subsection{Dipolar interaction of quantum rotors}

When quantum rotors get sufficiently close to each other, 
they can no longer be regarded
as being isolated,
thus, interaction between rotors should be taken into account.
Let us consider two rotors with dipolar moments $\bm{p}_i$ and $\bm{p}_j$
separated by a distance vector $\bm{R}$.
Expanding the interaction potential in terms of $1/R$,
the lowest-order approximation yields the dipolar term
\begin{equation}
W_{ij} = \frac{1}{4\pi\ep_r}\left\{
 \frac{\bm{p}_i\cdot\bm{p}_j}{R^3}
-3\frac{(\bm{p}_i\cdot\bm{R})(\bm{p}_j\cdot\bm{R})}{R^5}
\right\}.
\label{eq_05}
\end{equation}
Here $\ep_r$ is the dielectric constant of Ge crystals.
Discarding a numerical factor, 
the energy scale of dipolar-interaction potential (\ref{eq_05}) is characterized
by the quantity
\begin{equation}
J=\frac{p^2}{4\pi \ep_r R^3}.
\label{eq_06}
\end{equation}
For the case $E_K\gg J$, the rotors are nearly isolated 
so that the interaction potential (\ref{eq_05})
can be treated as a perturbation.
When $E_K\le J$, on the other hand, 
the quantum nature of coupled rotors is completely different
from that of isolated ones
because of their strong dipolar interaction.

At finite temperatures, the thermal fluctuation suppresses the dipolar interaction between the rotors.
Therefore, we introduce the upper cut-off length $R_{\rm max}$
defined by $J(R_{\rm max})\approx k_B T$,
or equivalently, 
\begin{equation}
R_{\rm max}=\left( \frac{p^2}{4\pi\ep_r k_B T} \right)^{\frac13}.
\end{equation}
Within this cut-off length, the two rotors correlate via dipolar interaction.
The value of $R_{\rm max}$ depends on the temperature $T$.
For example, $T=10$mK results in the cut-off length $R_{\rm max}\sim 30$\AA.
For actual oxygen concentration $\rho=10^{17}-10^{18}$ cm$^{-3}$,
this length is much less than the mean separation of adjacent rotors in Ge:O.
The length $R_{\rm max}$ further shortens with increasing temperatures.
Therefore, most rotors may be considered as non-interacting rotors
above the mK range.
However, it should be noted that the inhomogeneous distribution of rotors
makes the formation of clusters consisting of two or more rotors.
These clustered rotors yield peculiar low-energy excitations originating from
the dipolar interaction between rotors,
which significantly affect the properties of Ge:O at 1 K and below,
as will be shown in the subsequent sections.
Hereafter, we fix the cut-off length $R_{\rm max}= 30$ \AA\; for the temperature range
under consideration; for simplicity, we do not consider the discreteness of the lattice of Ge crystals.

Before closing this section,
the spatial correlation of impurity distribution should be mentioned.
We have assumed 
that no correlation exists in the distribution of oxygen impurities.
Yet this assumption is not always correct.
When the sample is prepared by pulling from the melt,
the correlation may be established at temperatures where oxygen impurities can diffuse.
Subsequently the correlated distribution of oxygen would be fixed on quenching.
It is difficult to definitely account for this correlation
since it depends essentially
on the history of the sample preparation.
Therefore, as a rule, we take a random distribution for oxygen 
with no spatial correlation,
bearing in mind that it is valid for the case of 
the absence of correlations.

%----------------------------------------------------------------------------------
\begin{figure}[ttt]
\hspace*{-1cm}
\includegraphics[width=10.5cm]{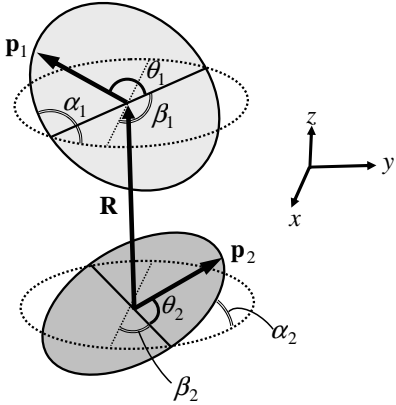}
\vspace*{-1cm}
\caption{Definitions of the coordinates of two quantum rotors 
carrying dipole moments $\bm{p}_1$ and $\bm{p}_2$.
The two rotors are separated by a vector $\bm{R}$ and interact with each other 
via the dipolar interaction potential $W_{12}$.}
\label{fig_02}
\end{figure}
%---------------------------------------------------------------------------------

\section{Low-energy excitations of paired rotors}

In this section, we focus mainly on the quantum character
of paired rotors ($n=2$).
This is because the behavior of clustered rotors with $n\ge 3$
is essentially understood by referring to those of paired rotors.
Emphasis will be placed on the fact that
strongly interacting rotors give rise to peculiar low-lying excitations;
the eigenenergies of these excitations are well below the energies
of first-excited states for an {\it isolated} rotor.
We will see that the low-lying excitations of paired rotors 
markedly contribute to low-temperature
properties of Ge:O below 1 K.

\subsection{Paired-rotor system}

Suppose that two quantum rotors with dipole moment $\bm{p}_i (i=1,2)$
are positionally separated by the vector $\bm{R}$ as depicted in Fig.~\ref{fig_02}.
The Hamiltonian for paired rotors is given by
\begin{equation}
H=K_1+K_2+V_1+V_2+W_{12},
\label{eq_08}
\end{equation}
where the definitions of $K_i$, $V_i$, and $W_{ij}$ are the same as those in
Eqs.~(\ref{eq_03}), (\ref{eq_04}), and (\ref{eq_05}),
respectively.
The explicit form of $W_{12}$ as a function of $(\th_1, \th_2)$
is given in Ref.~\cite{PRB}.

%------------------------------------------------------------------------------
\begin{figure}[ttb]
\vspace*{-2cm}
%\hspace*{-3.5cm}
\includegraphics[width=9.0cm]{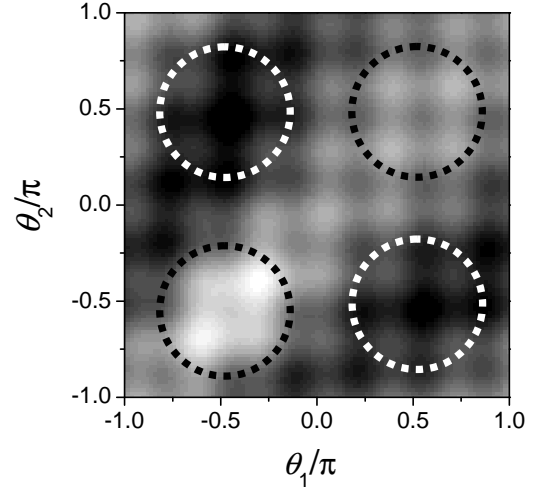}
\vspace*{-3.5cm}
\caption{Spatial profile of the potential field $F$ in the $(\th_1, \th_2)$ plane.
The dark regions correspond to negative values of the potential
$F$, while the white regions correspond to the positive.
Two global minima (maxima) are indicated by white (black) dashed circles.}
\label{fig_03}
\end{figure}
%----------------------------------------------------------------

As seen from Eq.~(\ref{eq_08}),
the quantum state of paired rotors is identified with that of single quantum particle
moving in a potential field
\begin{equation}
F(\th_1,\th_2)=V_1(\th_1)+V_2(\th_2)+W_{12}(\th_1,\th_2).
\label{eq_11}
\end{equation}
For later use, we plot the spatial profile of the potential $F$ in Fig.~\ref{fig_03}.
The angles $(\alpha_1, \alpha_2, \beta)$ defined in Fig.~\ref{fig_02}
are taken to be $(\pi/4, -\pi/6, \pi/3)$,
and the interaction energy $J$ is taken to be equal to 
the amplitude of the hindering potential.
The dark regions correspond to negative values of the potential
$F$, while the white regions correspond to the positive.
A spatial fluctuation with a period $\pi/3$
stems from the hindering term $V_i \propto \cos6\th_i$.
We notice that two global minima (maxima) appear at the anti-parallel (parallel)
dipolar configurations
as indicated by a white (black) dashed line.
These global minima and maxima indicate that 
two dipoles prefer to adopting an anti-parallel configuration.
The positions of two minima (and maxima) generally deviate from the lines
$|\th_2-\th_1|=0$ and $|\th_2-\th_1|=\pi$
due to the effect of the finite angles $(\alpha_1, \alpha_2, \beta)$.
For any set of angles $(\alpha_1, \alpha_2, \beta)$, 
however, the relative difference between the position of the minimum
and the maximum
are invariant with $\Delta\th_1=\Delta\th_2=\pi$.

\subsection{Eigenenergies for paired rotors}

The Schr\"odinger equation $H \Ps(\th_1, \th_2) = E \Ps(\th_1, \th_2)$
for the Hamiltonian (\ref{eq_08}) is solved by taking the solution in the form
\begin{equation}
\Ps(\th_1, \th_2) = \sum_{l_1, l_2} C_{l_1,l_2} \exp\{-i(l_1 \th_1 + l_2 \th_2)\},
\label{eq_12}
\end{equation}
where $l_i$ ($i=1,2$) takes an integer value $l_i=0,\pm1,\pm2,\pm3,\cdots$.
The coefficients $C_{l_1, l_2}$ are obtained numerically by diagonalizing 
the finite matrix of $H$ in the subset of basis states with
$|l_1|\le l_1^c$ and $|l_2|\le l_2^c$.
The cut-off values $l_1^c$ and $l_2^c$ are increased 
until the considered eigenvalues converge within the desired accuracy.
For actual calculations, we set $l_1^c=l_2^c=30$ in order to obtain precise low-energy states.

%------------------------------------------------------------------------
\begin{figure}[ttb]
\vspace*{-0.5cm}
\includegraphics[width=7.0cm]{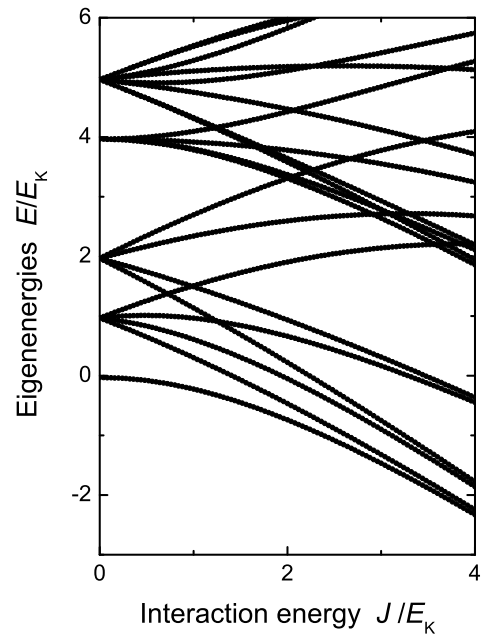}
\vspace*{-0.5cm}
\caption{The eigenenergy spectrum for paired rotors
as a function of the interaction energy $J$.
The relative angles of rotors are taken to be $(\alpha_1,\alpha_2,\beta)
= (\pi/4,-\pi/6,\pi/3)$, as an example.}
\label{fig_04}
\end{figure}
%-------------------------------------------------------------------------

Figure \ref{fig_04} shows the eigenenergy spectra for paired rotors
as a function of the dipolar interaction energy $J$.
Angles $(\alpha_1,\alpha_2,\beta)$ were chosen
to be $(\pi/4,-\pi/6,\pi/3)$ as an example.
When $J=0$, the eigenenergies are almost degenerate
at $E/E_K=l_1^2+l_2^2=0,1,2,4,5,8,\cdots$. 
An increase in $J$ leads to a split off of the degeneracies at $J=0$,
followed by a spread along the ordinate.
Note that the energy difference between the ground state and the first-excited state
decreases monotonously with increasing $J$.
For large $J\ge E_K$, these two eigenlevels form a doublet with very small energy splitting.
Besides, a part of higher eigenlevels also form doublets 
at $J\ge E_K$ as shown in Fig.~\ref{fig_04}.
As a consequence, a few doublets locate within the low energy region
$[E_G, E_G+E_K]$, where $E_G$ is the eigenenergy of the ground state.
Those doublets locating at the above energy region 
are what we call ``very-low-energy excitations" of paired rotors.
We notice that the very-low-energy excitations of paired rotors are generated
only when the rotors are strongly coupled.

The occurrence of the doublets for $J\ge E_K$ is attributed to 
the spatial profile of the potential field $F(\th_1,\th_2)$ defined in Eq.~(\ref{eq_11}).
For strongly paired rotors satisfying the condition $J\ge E_K$,
the depth of the two minima in the $\theta_1$-$\theta_2$ plane
become so large that the field $F$ eventually forms 
a double-well potential (See Fig.~\ref{fig_03}).
As a result, the behavior of paired rotors is described by a quantum particle
moving in a double-well potential whose barrier height is the order of the interaction energy $J$.
When the kinetic energy of the particle is much smaller than the barrier height,
eigenfunctions at low energies are described by a superposition of two wavefunctions
localized in the respective wells.
A slight overlap of the localized wavefunctions results in a doublet
with small energy splitting.
By raising the interaction energy $J$, therefore,
the barrier height becomes so large that the energy splitting monotonously decreases
as shown in Fig.~\ref{fig_04}.

\subsection{Low-energy excitations in clustered rotors}

%------------------------------------------------------------------------
\begin{figure}[ttb]
\vspace*{-1cm}
\includegraphics[width=7.7cm]{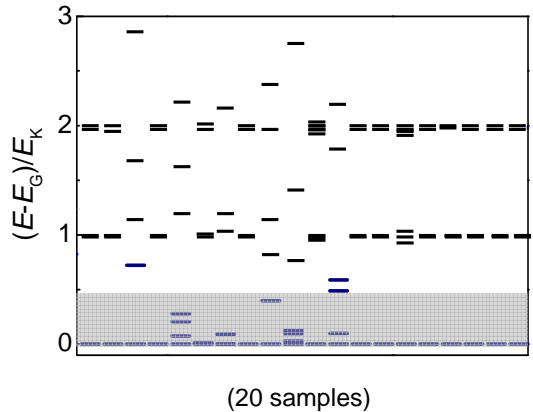}
\vspace*{-3.7cm}
\caption{Eigenenergy distributions for clustered rotors.
Twenty clusters were randomly chosen with various sizes 
and various spatial configurations.
}
\label{fig_05}
\end{figure}
%-------------------------------------------------------------------------

Doublets locating at very-low energies also occur in clustered rotors with larger sizes $n\ge 3$.
Figure \ref{fig_05} gives the energy spectra of the clustered rotors;
twenty clusters of various sizes and various spatial configurations were randomly chosen.
For any clusters, the ground state energies are fixed at zero
for the sake of illustration.
Most eigenlevels are located densely at specific energies
$E/E_K = 0, 1, 2, \cdots$, {\it i.e.},
at eigenenergies of an ensemble of isolated rotors.
This indicates that the rotor's interaction within individual clusters 
is not significantly large.
Nevertheless, a limited number of eigenlevels form doublets
showing energy splitting much smaller than $E_K$.
These doublets originate from the small separation $R$ of the rotors
involved in each cluster.
Namely, very-low-energy excitations in clustered rotors
can be found in clusters of any size because of a statistical fluctuation of
the rotor's spatial distribution.

Recall that the value of the kinetic energy $E_K$ for an actual Ge$_2$O unit
equals $2.3$ K.
This implies that only the very-low-energy excitations 
satisfying the condition $E/E_K<0.5$
are relevant to the properties of Ge:O below 1 K.
Further, when considering the properties of Ge:O far below 1 K,
the presence of doublets with the {\it lowest} energies would be especially important.
In the next subsection,
we discuss the energy distribution of such the lowest-energy excitations,
which would contribute to the specific heat of Ge:O far below 1 K.

%-----------------------------------------
%
\subsection{Two-level tunneling systems in Ge:O}

As mentioned before,
the occurrence of peculiar very-low-energy excitations stems from
the potential field $F(\th_1, \th_2)$ defined in Eq.~(\ref{eq_11}).
For strongly coupled rotors, the field $F$ forms a double-well potential
in the $\th_1$-$\th_2$ plane,
and the quantum behavior of the rotors is described by a quantum particle 
moving in the double-well potential.
Thereby, strongly coupled rotors
yields locally a quantum tunneling system.
Those tunneling systems are randomly distributed in Ge:O
due to an inhomogeneous distribution of oxygen impurities.
The height of the potential barrier $U$ separating the two potential minima
depends on the distance $R$ between rotors
as well as the internal configuration of interacting rotors.
Discarding numerical factors, the energy scale of the barrier height $U$
is estimated of the order of the interaction energy $J$.

The existence of tunneling systems in crystalline solids has also been demonstrated
in alkali halides with substitutional impurities such as KCl:Li, NaCl:OH, and KBr:CN.\cite{Wurger}
In these solids, tunneling occurs due to a substitutional single defect sitting
in sites offered by the host lattice. 
Each defect has an electric or elastic dipole moment. 
The possible orientation of the dipole is determined 
by the local potential minima in crystalline environment 
surrounding the defect. 
As a result, the defect atom can tunnel through several equivalent potential minima.

Regarding KBr:CN, Solf and Klein \cite{Solf_Klein} have theoretically investigated
the energy spectra of the system with paired tunneling dipoles.
Polar molecules of CN$^-$ dissolved in KBr form 
interacting dipoles oriented in eight directions.
For 340 ppm of CN$^-$ in KBr, they showed that the distribution function $P(\ep)$
of the lowest excitation energy $\ep$ for paired {\it elastic} dipoles
changes from $P(\ep)\sim {\rm const.}$ for $\ep/k_B<$ 100 mK
to $P(\ep)\propto \ep^{-1/2}$ for $\ep/k_B<$ 10 mK. \cite{Solf_Klein}
This yields a power-law temperature dependence of the specific heats
as $C(T)\propto T$ at 10 mK$<T<$1 K and $C(T)\propto T^{1/2}$ at 1 mK $<T<$ 10 mK.
The point is that paired tunneling impurities
engender novel low-energy excitations with small energy splitting $\ep$.
This is actually realized for interacting quantum rotors in Ge:O
as mentioned earlier,
though the explicit form of $P(\ep)$ remain unclear.
The microscopic character of interacting constituents in Ge:O
(weakly hindered quantum rotors)
is evidently distinct from that in alkali halides ($n$-oriented tunneling dipole).
Nevertheless, the difference is not influential to the emergence of the low-energy excitations
with small energy splittings.

Hereafter, we discuss the energy dependence of the distribution function 
$P(\ep)$ of the lowest energy excitations $\ep$
for strongly paired rotors in line with the theory of two-level tunneling systems.
\cite{AHV,Phillips}
The height of the tunneling barrier $U$ between two potential minima
scales as $U\sim J \propto R^{-3}$.
Thereby the WKB approximation \cite{Landau} allows us to evaluate the tunneling amplitude $\Delta_p$
expressing the coupling between two localized wavefunctions as
\begin{equation}
\Delta_p \sim \gamma \exp\left(-a U^{1/2} \right) \sim \gamma \exp\left(-aR^{-3/2}\right).
\label{wkb_1}
\end{equation}
The distribution function $P(\ep)$ is given by
the following three dimensional integral:
\begin{equation}
P(\ep) \sim \int d\bm{R}\; \delta\left\{ \ep - \Delta_p(R) \right\}.
\label{wkb_2}
\end{equation}
Substituting Eq.~(\ref{wkb_1}) into Eq.~(\ref{wkb_2}), 
we have
\begin{equation}
P(\ep) \propto \frac{1}{\ep \left\{ \log(\gamma/\ep)\right\}^3}.
\label{eq_12new}
\end{equation}
For strongly coupled rotors, the energy splitting $\ep$
is much smaller than the parameter $\gamma$.
When $\ep\ll \gamma$,
the function $P(\ep)$ given in Eq.~(\ref{eq_12new}) rapidly increases with decreasing $\ep$.
Its asymptotic behavior is approximated by the power law of $P(\ep)\propto \ep^{-\alpha}$,
where the exponent $\alpha$ is of the order of unity or less.

The ensemble of the two-level tunneling systems that obey the distribution function $P(\ep)$
contributes to the specific heat $C(T)$ by the following formula: \cite{AHV,Phillips}
\begin{equation}
C(T) = \int d\ep P(\ep) C_0 (\ep,T),
\end{equation}
and
\begin{equation}
C_0(\ep,T) \propto \frac{1}{k_B T^2}
\mbox{\rm sech}^2 \left( \frac{\ep}{2k_B T} \right). \label{eq_14new}
\end{equation}
Hence, the asymptotic behavior of $P(\ep)\propto \ep^{-\alpha}$
results in the power-law temperature dependence of the specific heats
$C(T)\propto T^{1-\alpha}$.
This implies that the specific heats for Ge:O, as in the case of KBr:CN,
should exhibit power-law behavior at sufficiently low temperatures.

It should be noted that the exponent of the predicted specific heat 
$C(T)\propto T^{1-\alpha}$ for Ge:O does not need to coincide with 
the exponent of the power-law specific heat for KBr:CN.
This is due to the difference of the microscopic structure of tunneling constituents
between in the two systems.
For KBr:CN, it has been assumed \cite{Solf_Klein} that the local potential minima
that determines the possible orientation of dipoles 
are invariant to the increase in the interaction energy $J$.
This assumption has led to the consequence that
the lowest excitation energies $\ep$ are proportional to $J^{-1}$ and $J^{-2}$,
within the second- and third-order perturbation theory, respectively.
As a result, the $\ep$-dependence of the density of states $P(\ep)$ can be solved analytically,
which leads to a linear or $T^{1/2}$ specific heats depending on the temperature range.
In contrast, for interacting quantum rotors in Ge:O,
the energitically-preferable orientation of rotors
complicatedly depend on both $J$ and the spatial configuration 
of interacting rotors, namely, by the angles ($\alpha_1,\alpha_2,\beta$).
Hence, the lowest excitation energies $\ep$ can not be expressed 
as a simple function of $J$ and ($\alpha_1,\alpha_2,\beta$),
which prevent us from obtaining analytically an explicit form of $P(\ep)$ and $C(T)$.
In order to clarify the $T$-dependence of the low-temperature specific heats $C(T)$
in Ge:O, therefore, it is crucial to numerically simulate $C(T)$ directly.

%--------------------------------------------
\begin{figure*}[ttb]
\begin{center}
\includegraphics[width=17.0cm]{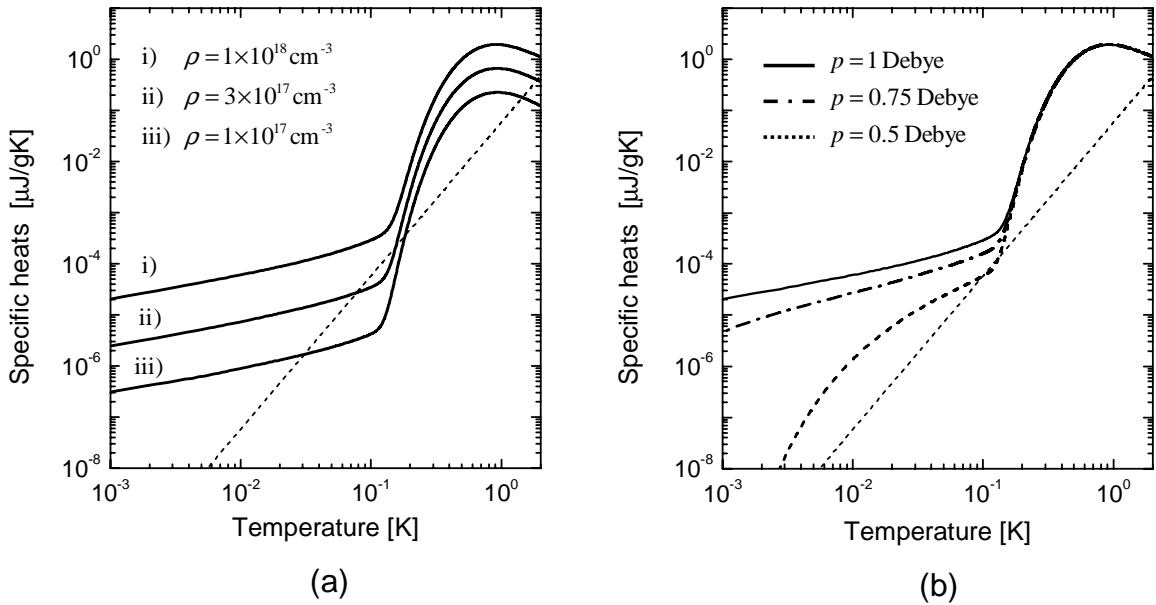}
\end{center}
\vspace*{-0.5cm}
\caption{(a) Calculated results of specific heats $C(T)$ for Ge:O.
The concentration of rotors $\rho$ is varied as follows:
(i) $\rho=10^{18}$ cm$^{-3}$, (ii) $3 \times 10^{17}$ cm$^{-3}$,
(iii) $\rho=10^{17}$ cm$^{-3}$, with the dipole moment fixed at $p=1$ Debye.
The dashed-dotted line follows 
the Debye specific heat for pure Ge crystals deduced from Ref.~\cite{Schnelle}.
(b) Results of specific heats $C(T)$ for $\rho=10^{18}$ cm$^{-3}$
with varying the definition of $p$;
(i) $p=1$ Debye, (ii) 0.75 Debye, and (iii) 0.5 Debye.}
\label{fig_06}
\end{figure*}
%----------------------------------------------------------------
%

%%%%%%%%%%%%%%%%%%%%%%%%%%%%%%%%%%%%%%%%%%%%
%
\section {Specific Heats}
%
%%%%%%%%%%%%%%%%%%%%%%%%%%%%%%%%%%%%%%%%%%%%

\subsection{Method}

Now let us consider the specific heat for a Ge:O  sample 
having an impurity concentration $\rho$.
This sample contains a large number of quantum rotors;
most rotors are isolated because of large separations,
while a part of neighboring rotors establish clusters having various sizes $n$.
The specific heat $C_{\rm cls}$ for an individual cluster
is calculated straightforwardly 
through its partition function $Z_{\rm cls}=tr(e^{-\beta H})$
and internal energy $U_{\rm cls}=-\frac{\partial}{\partial \beta} \log Z_{\rm cls}$,
\begin{equation}
C_{\rm cls} = \frac{dU_{\rm cls}}{dT}.
\label{eq_15}
\end{equation}
When calculating specific heats for a whole Ge:O sample,
on the other hand, we have to average the contributions of all clusters involved in the system:
\begin{equation}
C(T) = \frac1V \sum_{{\rm cls}}C_{\rm cls}.
\label{eq_16}
\end{equation}
Here $V$ represents the volume of the system to be considered.
Equation (\ref{eq_16}) amounts to considering all clusters with different sizes 
and with different spatial configurations.
This thereby gives the specific heat for a Ge:O sample, namely, 
for a particular realization of the rotor's spatial distribution.
Since this distribution strongly depends on the samples used,
we have to take sample averages in order to obtain 
the specific heat for a typical Ge:O sample.

To perform the above procedure,
we have numerically simulated more than 100 samples of Ge:O;
in each sample, 10$^6$ rotors are inhomogeneously distributed in a cubic system,
then are factorized into many $n$-sized clusters.
The volume of the system is defined by $10^6/\rho$,
and the lower cut-off $R_{\rm min}=3$ \AA$\;$ is introduced
in order to prevent two (or more) oxygen impurities from getting closer 
than the bond length of Ge-O-Ge.

\subsection{Numerical results}

Figure \ref{fig_06}(a) shows the numerical results of specific heat $C(T)$ 
for various concentrations of rotors $\rho$:
(i) $\rho=10^{18}$ cm$^{-3}$, (ii) $3 \times 10^{17}$ cm$^{-3}$,
and (iii) $10^{17}$ cm$^{-3}$.
The dipole moment $p$ of quantum rotors is set to be 1 Debye
as estimated before.
We also plotted the Debye specific heat $C(T)=aT^3$
with $a=5.83 \times 10^{-2}$ $\mu$J g$^{-1}$ K$^{-4}$.
The value of $a$ is evaluated based on the experimental data
observed at the temperature range 2.8-100 K \;\cite{Schnelle}.
In Fig.~\ref{fig_06} (a), broad peaks at around $T=1$ K commonly appear
for all rotor concentrations.
These peaks are caused by the fact that most rotors contained in the samples
are independent of each other.
For isolated rotors, the level-splitting between
the ground state and the first-excited ones
is equal to $E_K = 2.3$ K as seen in Fig.~\ref{fig_04}.
Thus the ensemble of isolated rotors contributes to specific heats
as the Schottky peaks at around $E_K/2 \sim 1$ K.

Of particular interest is 
the temperature dependence of specific heats $C(T)$ below the Schottky peaks.
At temperatures $T<0.1$ K, the calculated results obey
a power-law temperature dependence of the form 
$\propto T^{0.5}$.
The exponent $0.5$ seems to be independent of the impurity concentration $\rho$,
whereas the magnitude of $C(T)$ shifts downward
with decreasing $\rho$.
Furthermore, the magnitudes of $C(T)$ for temperatures $T < 0.1$ K
are proportional to $\rho^2$,
while those for $T > 0.1$ K (around the Shottky peak)
show a trivial linear dependence on $\rho$.
Interestingly, the same $\rho^2$-dependence of the power-law specific heats
has been pointed out in KBr:CN \;\cite{Solf_Klein}.
The physical origin of these peculiar behaviors in $C(T)$
will be addressed soon below.

We have also clarified the dependence of $C(T)$ on the definition of 
the value of $p$ for individual rotors.
Figure \ref{fig_06} (b) shows the resulting $C(T)$ for $p=1$ Debye (solid line),
$0.75$ Debye (dashed-dotted), and $0.5$ Debye (dashed).
Decreasing the value of $p$ reduces the magnitude of $C(T)$,
and results in a deviation of $C(T)$ from the power-law temperature dependence.
Nevertheless, 
the considerably excess specific heat below 0.1 K remains observed for all $p$s,
which is a manifestation 
of the dipole interaction of rotors consisting of clusters.
It should be emphasized that, for all numerical conditions in Figs.~\ref{fig_06} (a) 
and \ref{fig_06} (b),
the results of $C(T)$ below 0.1 K well overcome
the Debye specific heat.
This demonstrates that our theoretical results for specific heats, 
both the Schottky peaks and the power-law behavior,
can be observed in experiments.

\subsection{Two anomalies in the excess specific heats below 0.1 K}

In Figs.~\ref{fig_06} (a), 
we have found two striking features in the excess specific heats below 0.1 K.
The one is the power-law temperature dependence of $C(T)\propto T^{0.5}$,
and the other is the $\rho^2$-dependence of the magnitude of $C(T)$.
To reveal the physical origin of the two features,
we have clarified the contribution of the ensembles of clustered rotors of respective sizes.

\subsubsection{Power-law specific heats}

\begin{figure}[ttb]
\vspace*{-1.0cm}
\hspace*{-0.5cm}
\includegraphics[width=9.5cm]{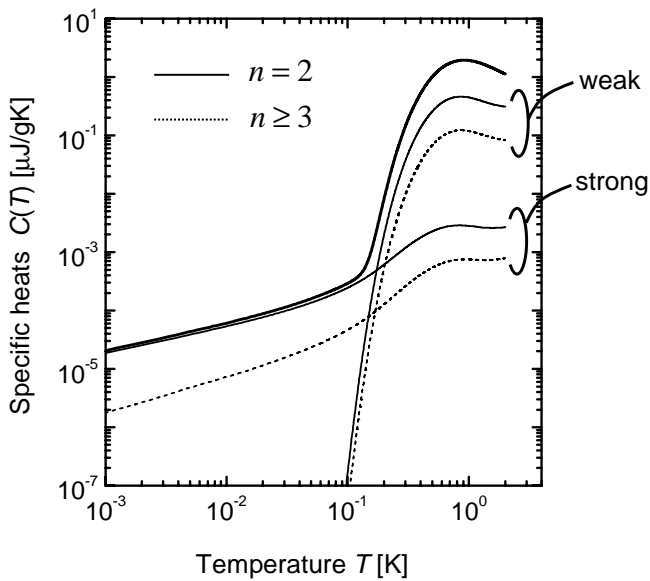}
\vspace*{-4.0cm}
\caption{The result of $C(T)$ for $\rho=10^{18}$ cm$^{-3}$ and $p=1$ Debye (thick line)
is replotted, together with four components stemming from ensembles of 
paired ($n=2$) and clustered ($n=3$) rotors.
The meanings of the term ``weak"  and ``strong" are given in text.}
\label{fig_07}
\end{figure}

\begin{figure}[ttb]
\vspace*{0cm}
\includegraphics[width=8.8cm]{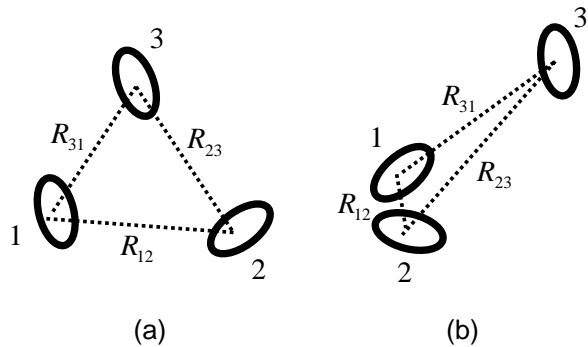}
\vspace*{-6.0cm}
\caption{Schematic illustrations of rotor clusters with $n=3$:
(a) A cluster formed by weakly interacting rotors.
All separations $R_{ij}$ exceed the characteristic length $R_c=10$\AA.
(b) A cluster established by strongly interacting rotors.
The separation $R_{12}$ is smaller than $R_c$.}
\label{fig_08}
\end{figure}

In Fig.~\ref{fig_07}, the result of $C(T)$ for $\rho=10^{18}$ cm$^{-3}$ and $p=1$ Debye
is replotted by a thick line.
In addition, we exhibit four components consisting of $C(T)$
for a complete Ge:O system;
solid thin lines show the contribution of the ensemble of coupled rotors ($n=2$),
and dashed lines show that of clustered rotors ($n\ge 3$), as denoted in the figure.
The two components, indicated by the term ``weak", correspond to the ensemble of clusters
established by weakly-interacting rotors;
within those individual clusters, all rotor separations are larger than 
the characteristic value $R_c=10$\AA$\;$ (See Fig.~\ref{fig_08} (a)).
Because of the large separations,
the eigenenergies of those clusters are nearly 
equal to the eigenenergies of isolated rotors,
thus giving rise to the Schottky behavior appearing in Fig.~\ref{fig_07}.
On the other hand,
the other two components, indicated by ``strong", 
represent the contribution of the ensemble of clusters
including strongly-interacting rotors;
within those clusters, two or more rotors are closer than $R_c$ (See Fig.~\ref{fig_08} (b)).
As clearly seen from Fig.~\ref{fig_07}, the excess specific heats below 0.1 K
originate from the ensemble of strongly coupled rotors ($n=2$) with separations $R<R_c$.
Consequently, we conclude that the power-law behavior of $C(T)\propto T^{0.5}$
for a Ge:O sample originate from the very-low-energy excitations of strongly coupled rotors.

%---------------------------------------------------------------------------------
%----------------------------------------------------------------

\begin{figure}[ttb]
\vspace*{-0.0cm}
\includegraphics[width=8.5cm]{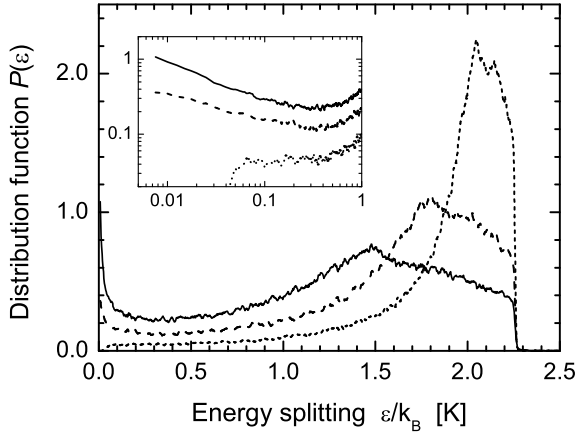}
\vspace*{-0.0cm}
\caption{Distribution functions of the energy splitting $\ep$
between the ground state and the first-excited one.
The definition of $p$ is set to be $p=1$ Debye for solid line,
0.75 Debye for dashed line, and 0.5 Debye for dotted line.
}
\label{fig_09}
\end{figure}

Our conclusion mentioned above is justified by evaluating the behavior of $P(\ep)$
that directly relates to the low-temperature specific heats through Eq.~(\ref{eq_14new}).
Figure \ref{fig_09} shows the distribution function $P(\ep)$ 
of the energy splitting $\ep$ between the ground state and the first-excited ones
for strongly coupled rotors;
the definition of $p$ is set to be $p=1$ Debye for solid line,
0.75 Debye for dashed line, and 0.5 Debye for dotted line.
All numerical results are normalized as $\int P(\ep)d\ep = 1$.
The definitions $R_{\rm min}=3$\AA$\;$ and $R_c=10$\AA$\;$ give
a lower limit for the energy splitting $\ep_{\rm min}$ 
and an upper one $\ep_{\rm max}$, respectively.
Numerical results of $P(\ep)$ show a maximum at around 1.5-2.0 K
followed by a rapid decrease with decreasing $\ep$.
Interestingly, for a lower $\ep$ than 0.1 K,
the function $P(\ep)$ (except for dotted line) obey the power-law form 
$P(\ep) \propto \ep^{-\alpha}$ within a range over one order
(See the inset of Fig.~\ref{fig_09}).
This resulting power-law is consistent with the theoretical consequence given in Sec.~III-D.
In the case of $p=1$ Debye, in particular, the exponent $\alpha$ takes 0.5;
this reproduces the power-law specific heat $C(T)\propto T^{0.5}$
for sufficiently low temperatures $T\ll \ep_{\rm max}/k_B$,
which is clearly demonstrated in Fig.~\ref{fig_06} (a).

We have numerically confirmed that the exponent $0.5$ in the power-law form of $P(\ep)\propto \ep^{-0.5}$
is invariant to the change in the rotor concentration $\rho$ under the condition $p=1$ Debye.
On the other hand, for smaller $p$,
the exponent $\alpha$ of $P(\ep)\propto \ep^{-\alpha}$ for $\ep/k_B \ll 0.1$ K gradually decreases,
and finally $P(\ep)$ becomes a constant 
as shown by the dotted line (for $p=0.5$ Debye) in the inset of Fig.~\ref{fig_09}.
This results from the fact that, when $p$ decreases, the lower bound $\ep_{\rm min}$
increases because of the reduction of the interaction energy $J$.
Eventually $\ep_{\rm min}$ gets larger than the energy range where the function $P(\ep)$ clearly
exhibits a power-law behavior $P(\ep)\propto \ep^{-\alpha}$ with positive $\alpha$.
As a result, $P(\ep)$ seems to be constant for the condition of $p=0.5$ Debye.
In addition, the energy range where $P(\ep)$ is constant
is rather narrow;
this is the reason why the specific heats for the condition of $p=0.5$ Debye exhibit
a power-law temperature dependence only within a rather narrow temperature range
as shown in Fig.~\ref{fig_06} (b).

%---------------------------------------------------------------------------------
%----------------------------------------------------------------

%--------------------------
\subsubsection{$\rho^2$-dependence of excess specific heats}

Next we turn to the $\rho^2$-dependence of the magnitude of $C(T)$ below 0.1 K.
The $\rho^2$-dependence of the low-temperature specific heats
has also been pointed out for KBr:CN as a consequence of the virial expansion
of the free energy up to $\rho^2$ \;\cite{Solf_Klein}.
In the following, we derive the $\rho^2$-dependence of $C(T)$ straightforwardly
by considering the number of strongly coupled rotors $N_p$ in the system.
The quantity $N_p$ can be expressed by the relation $N_p=N\times u$.
Here, $N$ is the total number of rotors in the system,
which is apparently proportional to the rotor concentration $\rho$.
The quantity $u$ represents
the possibility that a given rotor finds another rotor
with a separation $R<R_c$.
Thereby $u$ reads
\begin{equation}
u = \int_{R_{\rm min}}^{R_c} P(R) dR,
\label{eq_19}
\end{equation}
where $P(R)$ is the distribution function for the separation $R$.

The explicit form of the function $P(R)dR$ is obtained as follows \cite{Mehta}.
$P(R)$ represents the probability of finding no rotor within a sphere of radius $R$
and one rotor in the spherical shell of the thickness $dR$, when one rotor is located at the origin.
In order to calculate $P(R)$,
we divide the sphere of radius $R$ into $m$ shells of thickness $R/m$.
Since the rotors are randomly distributed,
the probability of not finding a rotor in a sphere of radius $R$
is expressed by the probabilities of none of these $m$ shells containing rotors.
Thus, we can express the probability density $P(R)$ for large $m$
\begin{eqnarray}
\hspace*{-20pt}& & P(R)dR
= \left[1-\frac43 \pi\rho\left(\frac{R}{m}\right)^3\right] \nonumber \\
\hspace*{-20pt} &\times&
\Pi_{j=0}^{m-1} \left[1-4\pi\rho\left(\frac{jR}{m}\right)^2 \cdot \frac{R}{m} \right]
\times 4\pi \rho R^2 dR.
\label{eq_pr1}
\end{eqnarray}
Taking the limit $m\to \infty$, the probability of 
finding a nearest-neighbor rotor between $R$ and $R+dR$ is given by
\begin{equation}
P(R)dR=4\pi \rho R^2 \exp\left(-\frac43 \pi \rho R^3\right)dR.
\label{eq_pr2}
\end{equation}
The probability density $P(R)$ has its maximum at $R_0= (2\pi \rho)^{-1/3}$.
Since $R_c=10$\AA$\;$ is much smaller than $R_0$,
the function $P(R)$ can be approximated by $P(R)\approx 4\pi \rho R^2$.
Substituting it into Eq.~(\ref{eq_19}), we obtain the relation $u \propto \rho$.
As a result, the number of strongly coupled rotors $N_p$ is proportional to $\rho^2$.
This naturally leads to the peculiar $\rho^2$-dependence of the magnitude of $C(T)$ below 0.1 K,
as demonstrated in Fig.~\ref{fig_06} (a).

%---------------------------------------------------------------------------------
%----------------------------------------------------------------
%

Let us summarize the findings for low-temperature specific heats in Ge:O systems.
Numerical simulations have suggested the three anomalous behaviors in specific heats of Ge:O systems:
(i) the Schottky peaks at around 1 K,
(ii) the power-law temperature dependence below 0.1 K,
and (iii) the $\rho^2$-dependence of the magnitude of $C(T)$ below 0.1 K.
The first can be attributed to the ensemble of isolated rotors in the system,
while the latter two stem from the presence of strongly coupled rotors
with small separations $R<R_c=10$\AA.
Furthermore, the physical origin of the latter two anomalies
has been quantitatively understood through the distribution function of the energy splitting $P(\ep)$
as well as that of separation $P(R)$.
It is expected that our predictions will shed light on experimental research 
regarding the low-temperature properties of oxygen-doped semiconductors.

%%%%%%%%%%%%%%%%%%%%%%%%%%%%%%%%%%%%%%%%%%%%
%
\section{Dielectric Susceptibility}
%
%%%%%%%%%%%%%%%%%%%%%%%%%%%%%%%%%%%%%%%%%%%%

We now focus on the dielectric response for Ge:O at low temperatures.
The quantum tunneling dipoles in alkali halides crystals are known to
significantly contribute to the dielectric susceptibility
at low temperatures.
In a series of papers \cite{Klein}, 
Klein has theoretically investigated various dynamic and static quantities in terms of
the virial expansion,
suggesting that the low-$T$ dielectric susceptibility $\chi(T)$
contains a $-\log T$ term.
The experimental results of $\chi(T)$ for KCl:Li have exhibited an interesting dependence on
the impurity concentration $\rho$ and temperature $T$.
At $\rho$ of a few ppm, the susceptibility $\chi(T)$ monotonically increases with decreasing $T$,
and then approaches a constant value for $T \to 0$.
On the other hand, for the samples of 60 ppm or more,
$\chi(T)$ is no longer a monotonic function of $T$.
The susceptibility $\chi(T)$ displays a maximum at a temperature of approximately 300 mK,
and the magnitude of the maximum is pronounced by an increase in the impurity concentration $\rho$.
Further, an additional increase in $\rho$ of 1100 ppm surprisingly
results in a decreasing susceptibility,
which is a manifestation of the nonlinearity of the low-$T$ susceptibility $\chi(T)$
on $\rho$.
In Refs.~\cite{Enss,Wurger_Euro},
the occurrence of the maximum in $\chi(T)$ as well as the abovementioned non-linearity of $\chi(T)$ on $\rho$
is described by the theory
based on Mori's reduction method \cite{Wurger1,Wurger2}.

In this section, we examine the effect of the presence of the interacting quantum rotors 
on the dielectric susceptibility
$\chi(T)$ in Ge:O for an oxygen concentration $\rho=10^{17}-10^{18}$ cm$^{-3}$.
The maximum of $\chi(T)$ at 1 K, similar to that observed in KCl:Li, is also observed for Ge:O.
Interestingly, the origin of the maximum in $\chi(T)$ is directly attributed
to the presence of the low-lying excitations
of the interacting quantum rotors,
which is different from the interpretation of the maximum of $\chi(T)$ 
for KCl:Li, given in Refs.~\cite{Enss,Wurger_Euro}.

%----------------------
\subsection{Method}

The linear dielectric function for an individual cluster is described by the time-dependent
response function with respect to an external field,
\begin{equation}
\chi_{\mu\nu}^{\rm cls}(t-t') = \frac{i}{\hbar\ep} 
\langle \left[ P_{\mu}(t), P_{\nu}(t') \right] \rangle
\Theta(t-t').
\label{eq_20}
\end{equation}
Here, the angular brackets indicate that the thermal average
$\langle \cdots \rangle = Z^{-1} tr\left(\cdots e^{-\beta H}\right)$ is taken.
The operator $P_{\mu}(t)$ represents the $\mu$-component of the total dipole moment for the cluster,
and $\Theta(t)$ is the Heviside step function.
Since experiments provide quantities depending on frequency rather than on time,
we take the Fourier transform of Eq.~(\ref{eq_20}),
\begin{eqnarray}
\chi_{\mu\nu}^{\rm cls}(\omega,T) &=&
-\frac{2}{\ep Z}\sum_{j,l\ne j}
\langle E_j|P_{\mu}|E_l \rangle \langle E_l|P_{\nu}|E_j \rangle\nonumber \\
%\left| \langle E_j|p_{\mu}|E_l \rangle \right|^2 \nonumber \\
&\times&
 \frac{E_j-E_l}{(E_j-E_l)^2-(\hbar\omega)^2} \exp\left(-\frac{E_j}{k_B T}\right).
\label{eq_21}
\end{eqnarray}
Here, $|E_j \rangle$ means the eigenstates of the clustered rotors
belonging to the eigenenergy $E_j$.
We focus on the dielectric response of Ge:O$_i$ at frequencies $\omega$
lower than the kHz range;
in this case, the energy difference $|E_j-E_l|$ for arbitrary eigenstates
is larger than $\hbar \omega$ by many orders of magnitudes.
Thus in the sequel we take the zero-frequency limit of Eq.~(\ref{eq_21}).

For calculating the susceptibility $\chi_{\mu\nu}$ for a whole Ge:O$_i$ sample,
we have summarized the contribution of all clusters,
\begin{equation}
\chi_{\mu\nu}(T) = \sum_{\rm cls} \chi_{\mu\nu}^{\rm cls}(T),
\label{eq_22}
\end{equation}
and performed the sample average
as in the cases for calculating the specific heat.
In addition, the isotropy of Ge:O$_i$ allows us to consider
the quantity $\chi\equiv (\chi_{xx} + \chi_{yy} + \chi_{zz})/3$.
Consequently, the objective is the following quantity,
\begin{eqnarray}
\chi(T) &=& \sum_{\rm cls} \chi^{\rm cls}(T), \label{eq_23} \\
\chi^{\rm cls}(T) &=& -\frac{2}{3\ep Z} \sum_{j,l\ne j} 
\frac{|\langle E_j | P_{\mu} | E_l \rangle|^2}{E_j-E_l} 
\exp\left(-\frac{E_j}{k_{\rm B} T}\right),
\label{eq_24}
\end{eqnarray}
with arbitral component $P_{\mu}$.

%---------------------------------------------------------------------------------
\begin{figure}[ttb]
\includegraphics[width=8.5cm]{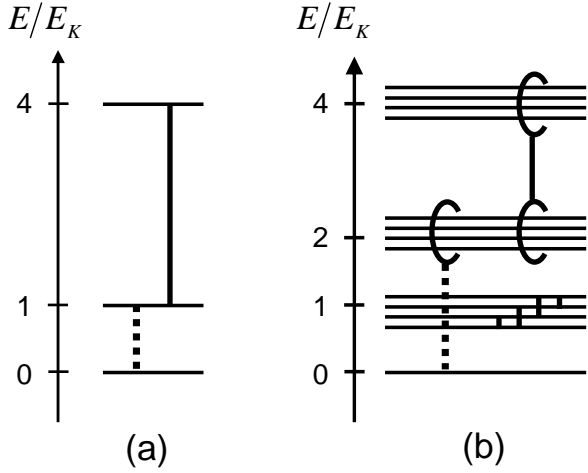}
\vspace*{0mm}
\caption{Schematic illustration of allowed dipolar transitions:
(a) the case for an isolated rotor, and (b) single paired rotors.
Vertical lines indicate the allowed dipolar transitions;
a dashed line refers to the allowed transition 
involving the ground state,
and a solid line refers to the remainder.}
\vspace*{0.0cm}
\label{fig_10}
\end{figure}
%----------------------------------------------------------------

\subsection{Contribution of isolated and paired rotors}

For later use, we examine the dielectric properties
of a isolated quantum rotor and that of paired rotors.
Figure \ref{fig_10} displays the low-lying energy levels of (a) an isolated rotor, and 
(b) weakly coupled rotors.
Vertical lines indicate the allowed dipolar transitions;
dashed lines refer to the allowed transition involving the ground state,
and solid lines to the remainder.
Dipolar transitions involving higher levels than $E/E_K=4$ are not shown here,
since those higher levels are only negligibly excited below a few degrees Kelvin.

For an isolated rotor, only two dipolar transitions contribute to the dielectric susceptibility
$\chi(T)$ below a few Kelvin.
On the other hand, for single paired rotors,
many eigenlevels lower than $4E_K$ contribute to the dielectric susceptibility
as shown in Fig.~\ref{fig_10} (b);
rings bundling four eigenlevels indicate that
the corresponding vertical line connects all of the four eigenlevels bundled.
As a consequence, 24 dipolar transitions are exhibited in Fig.~\ref{fig_10} (b).
Particularly important is the occurrence of 
the allowed transition at around $E\sim E_K$ connecting nearly-degenerate eigenlevels.
The small energy differences $| E_j-E_l |$ of two nearly-degenerate eigenstates
$| E_j \rangle$ and $| E_l \rangle$ magnify the absolute value of the fraction
$\frac{|\langle E_j | P_{\mu} | E_l \rangle|^2}{E_j-E_l}$
appearing in Eq.~(\ref{eq_24}),
then leading to an anomalous maximum in $\chi(T)$ at the temperature $T\sim E_K/k_B$ 
\;\cite{PRB}.
We emphasize that such almost degenerate levels
at $E\sim E_K$ emerge only for weakly paired (and clustered) rotors.
Namely, the remarkable contribution of the dipolar transitions
associated with nearly-degenerate levels occurs only for interacting rotors.
This implies that the presence of paired and clustered rotors affects
the dielectric properties of Ge:O at low temperatures.

We have confirmed that the maximum in $\chi(T)$ appears also for clustered rotors
with sizes $n\ge 3$.
Note that, however, the magnitude of $\chi(T)$ for clustered rotors
is generally reduced with increasing the cluster size.
This is because the many-body interaction of rotors with random configurations
prevents them from orienting in the direction of an electric field.
Indeed, the magnitude of $\chi(T)$ for clusters $n=5$ become less than that for paired rotors
by two orders or more.
Furthermore, the possibility of the occurrence of clusters with large sizes ($n\ge 6$)
is extremely small.
These facts allow us to ignore the presence of clusters with $n\ge 6$
for considering the dielectric susceptibility $\chi(T)$ in Ge:O systems.

%---------------------------------------------------------------------------------
\begin{figure}[ttb]
\hspace*{-1.0cm}
\includegraphics[width=9.5cm]{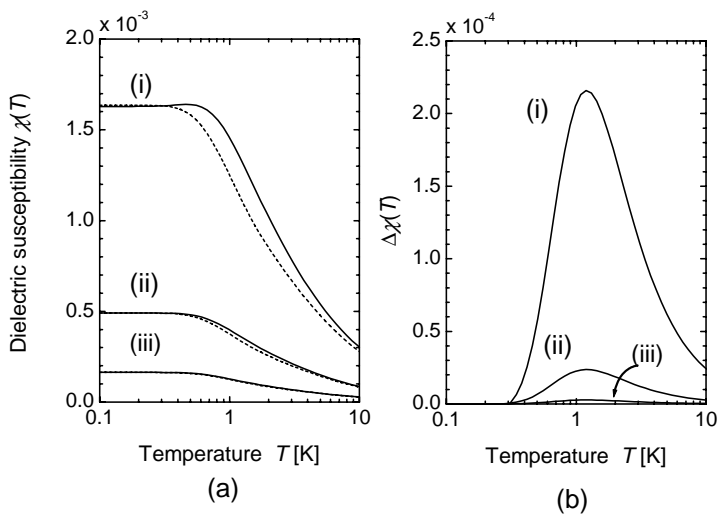}
\vspace*{-0.0cm}
\caption{
(a) The dc dielectric susceptibility $\chi(T)$ 
of the same four conditions shown in Fig.~\ref{fig_06}.
Solid lines show the total susceptibility $\chi(T)$,
and dashed lines show the contribution of the ensemble of isolated rotors.
Shallow maxima at around $T = 1.0$ K appear in all conditions.
(b) The differences $\Delta \chi(T)$ 
between the solid lines and the dashed lines exhibited in the figure (a).
}
\label{fig_11}
\end{figure}
%----------------------------------------------------------------

\subsection{Dielectric responses for Ge:O$_i$ }

Figure~\ref{fig_11} (a) shows the calculated results of the dc dielectric susceptibility $\chi(T)$
for Ge:O samples;
(i) $\rho=10^{18}$ cm$^{-3}$, (ii) $3 \times 10^{17}$ cm$^{-3}$,
and (iii) $10^{17}$ cm$^{-3}$ with fixing the dipole moment  $p=1$ Debye.
The solid lines show the total susceptibility $\chi(T)$,
while the dashed ones show the contribution of the ensemble of isolated rotors;
the latter can be analytically obtained from Eq.~(\ref{eq_24}),
since the eigenenergies and their eigenfunctions for isolated rotors
are analytically calculated.
The magnitude of $\chi(T)$ depends linearly on the concentration of oxygen $\rho$
and squarely on the dipole moment $p$,
which is simply understood from Eqs.~(\ref{eq_23}) and (\ref{eq_24}).
We find that an increase in $\rho$ gradually enhances
the maximum of $\chi(T)$ at about 1 K, which comes from
an increase in the number of paired and clustered rotors in the Ge:O sample \cite{foot_chi}.

To aid in illustration,
Fig.~\ref{fig_11} (b) exhibits the differences $\Delta \chi(T)$ 
between the solid lines and the dashed lines displayed in Fig.~\ref{fig_11} (a).
The maximum in $\Delta \chi(T)$ is located at about 1 K independent of the oxygen concentration.
Notably, when we compare the data for cases (i) and (ii),
we find that a change in oxygen concentration $\rho$ by a factor of {\it three}
enhances the quantity $\Delta \chi$ by almost {\it an order}.
Namely, the magnitude of $\Delta \chi$ depends almost squarely 
on the oxygen concentration $\rho$.
This $\rho^2$-dependence of $\Delta \chi$ can be explained using the same scenario as that for the power-law
specific heats (See Sec.~IV-C).
Consequently, the $\rho^2$-dependence of the quantity $\Delta \chi(T)$
as well as the maximum in $\Delta \chi(T)$
at about 1 K directly reflect the presence of paired and clustered rotors
in Ge:O systems.
We emphasize that the non-trivial $\rho^2$-dependence of $\Delta \chi$
as well as its maxima can be measured experimentally,
since the contribution of isolated rotors, depicted by the dashed line in Fig.~\ref{fig_11} (a),
has an analytic expression deduced from Eq.~(\ref{eq_24}).

It is worth noting that our results of $\chi(T)$, shown in Fig.~\ref{fig_11} (a),
are quite similar to the experimental data of $\chi(T)$ 
for KCl:Li observed at temperatures between 10 mK and 10 K \;\cite{Enss,Wurger_Euro}.
The experiments were performed
with Li concentrations ranging from 4 ppm to 1100 ppm,
revealing the maximum in $\chi(T)$ between 100 mK and 500 mK.
Further, the peak becomes more pronounced with increasing Li concentrations,
which is quite similar to the case of Ge:O as shown in Fig.~\ref{fig_11}.
However, despite the similarity, the physical origin of the maximum in $\chi(T)$
predicted for Ge:O is different from that observed for KCl:Li \;\cite{Enss,Wurger_Euro}.
In the latter case, the peak in $\chi(T)$ is identified with the relaxation peak
that indicates the collective motion of the tunneling dipoles \cite{Wurger1, Wurger2}.
On the other hand, the maximum in $\chi(T)$ for Ge:O systems
is not related to the relaxation process.
The maximum originates from the dipolar transition of
the nearly degenerate levels for weakly coupled rotors.
In other words, for the interacting dipoles, the maximum in $\chi(T)$ possibly occurs without
the relaxation process.
The presence of the allowed dipolar transition of the nearly degenerate levels is of importance.
Consequently, the maximum in $\chi(T)$ for Ge:O that we have observed
is different from the relaxation peak observed in KCl:Li.
When the oxygen concentration is further increased,
the collective motion of the rotors would become relevant.
In this case, the relaxation effect might manifest itself in the temperature dependence of $\chi(T)$,
which is similar to that in KCl:Li.
Further discussion are required to clarify this point.

%%%%%%%%%%%%%%%%%%%%%%%%%%%%%%%%%%%%%%%%%%%%
%
\section{Discussions}
%
%%%%%%%%%%%%%%%%%%%%%%%%%%%%%%%%%%%%%%%%%%%%

In this section, we focus on the relevance of our findings to other physical systems
that also exhibit low-temperature anomalies similar to those in Ge:O.

\subsection{The dipole gap}

It should be remarked the relevance of our model to the dipolar gap theory \cite{Bara, Efros}.
It is known that the long-range interaction of constituents in solids
generally leads to a remarkably reduction of the density of low-energy excitations.
This reduction stems from the requirements of the ground state stability 
relative to two-particle excitations and more complex ones for the system of local interacting centers.
In the case of dipolar interaction, the density of states $P(\ep)$ for low energies
tends to be zero logarithmically as \cite{Bara} $P(\ep) \propto 1/\log (\gamma/\ep)$,
where $\gamma$ is an appropriate constant.
This soft gap, called the dipole gap, manifests itself in a singularity of the specific heats
as $C(T) \propto T/\log(\gamma/k_{\rm B} T)$.
Kirkpatrick and Varma \cite{Kirk}
have suggested the result $C(T)\propto T^{3/2}$ in terms of the Monte Carlo
simulations.
These conclusions disagree with our result of $C(T)$ and $P(\ep)$
for coupled dipolar rotors.

We believe that the above disagreement in the behavior of $C(T)$
is due to the following reasons.
The first has to do with rather dilute concentrations of rotors in Ge:O systems.
The occurrence of the dipolar gap requires a large number of dipole moments interacting simultaneously.
This is hardly realized in Ge:O systems,
since in crystalline Ge there exists the upper limit of solubility of oxygen \cite{Clauws}.
Namely, a high concentration of rotors sufficient to realize the dipole gap
cannot be achieved as far as Ge:O systems are concerned.

The second reason is the difference of the nature between {\it classical} dipole moments
and {\it quantum} dipolar rotors.
The derivation of the dipole gap in Ref.~\cite{Bara}
was based on the classical motion of charged particles.
Differing from that, we have considered the quantum character of rotating Ge$_2$O units
carrying dipole moments; in addition, the rotation of the Ge$_2$O units
is restricted to within a two-dimensional plane perpendicular to the Ge-Ge axis.
Thus the argument based on the coupled rotor model surely yields 
the density of states $P(E)$ and the $T$-dependence of $C(T)$ 
distinct from those in Ref.~\cite{Bara}.
Independent of the relevance for Ge:O, nevertheless,
it is interesting whether or not the dipolar gap occurs in the quantum dipolar rotors
in sufficiently dense concentrations.
The scaling scheme for disordered spin systems \cite{Fisch} may help to address this issue.

\subsection{Two-level tunneling states in amorphous solids}

It is known that the power-law specific heats at low temperatures
have been observed in various kinds of amorphous materials \cite{Zeller}.
In those systems, the power-law specific heats have been explained by a phenomenological model
based on two-level tunneling states (TLSs) with an assumed constant density of states \cite{AHV,Phillips}.
The presence of TLSs in various amorphous solids has been confirmed by many experiments
\cite{Esqui}.
There is, however, as yet no microscopic derivation for the tunneling states
or for the constant density of states.
In addition, the theoretical result $C(T)\propto T$
differs somewhat from the observed behavior
$C(T) \propto T^{1+\delta}$ with small deviations  $\delta$
depending on materials.
Recently found evidences for dipole-dipole interactions between TLSs
draw attention to the necessity of modifying or generalizing the simple TLS picture 
\cite{Salvino,Burin,Osheroff,Ladieu}.

Of particular interest is that the power-law specific heat in Ge:O can be interpreted
in line with the scenario analogous to that of the tunneling model.
Emphasis should be made on, nevertheless, the two crucial differences 
between the coupled-rotor model and the standard TLS model as follows:
(i) The coupled-rotor model
gives the clear microscopic origin for double-well potentials randomly distributed in Ge:O systems;
this is apparently in contrast with the phenomenological TLS model applied to amorphous solids.
(ii) While the TLS model in amorphous solids requires 
an artificial distribution function for two-level tunneling states,
our model naturally yields the power-law distribution function
with no artificial condition.
Consequently, the clarity of the coupled-rotor model enables us to argue 
the low-temperature properties of Ge:O quantitatively,
giving the theoretical prediction of the power-law specific heat in Ge:O.
This is also the case for alkali haride crystals containing tunneling dipoles.

\section{Conclusion}

In conclusion,
we have theoretically investigated the effect of the dipolar interaction of oxygen defects 
on various physical properties in crystalline germanium.
Numerical simulations have revealed that 
the dipolar interaction of nearby oxygen impurities engenders non-trivial low-lying excitations,
causing anomalous behaviors for Ge:O systems with an oxygen concentration $10^{17-18}$cm$^{-3}$.
The following were the main findings for the low-temperature anomalies of Ge:O:
(i) the Schottky peak in specific heats $C(T)$ at around 1 K,
(ii) the power-law temperature dependence below 0.1 K,
(iii) the $\rho^2$-dependence of the magnitude of $C(T)$ below 0.1 K,
(iv) the anomalous hump in the dielectric susceptibility $\chi(T)$ at around 1 K,
and (v) the $\rho^2$-dependence of the magnitude of the deviation $\Delta \chi(T)$
from the susceptibility of isolated rotors.
All these behaviors originate from the presence of interacting quantum rotors in Ge:O systems,
and can be understood quantitatively
by considering the contribution of the ensemble of interacting rotors.
We have accounted for the occurrence of non-trivial low-lying excitations
in interacting rotors in line with the two-level tunneling theory.
The picture is based on quantum tunneling in a local double-well potential,
thus making clear the relevance of the present results to other systems involving tunneling constituents.
We hope that our findings shed light on experimental research regarding
the low-temperature properties of oxygen-doped semiconductors
as well as on disordered systems with interacting dipole moments.

\begin{acknowledgments}
This work was supported in part by a Grant--in--Aid for Scientific
Research from the Japan Ministry of Education, Science, Sports and Culture.
One of the authors (H.S) thanks the financial supports from 
the Foundation Advanced Technology Institute.
A part of numerical calculations were performed on the SR8000
of the Supercomputer Center, Institute of Solid State Physics, 
University of Tokyo.
\end{acknowledgments}

%%%%%%%%%%%%%%%%%%%%%%%%%%%%%%%%%%%%%%%%%%%%
%
%     References
%
%%%%%%%%%%%%%%%%%%%%%%%%%%%%%%%%%%%%%%%%%%%%


\begin{thebibliography}{99}

\bibitem{Bosomworth} W.~Hayes and D.~R.~Bosomworth, Phys.~Rev.~Lett. {\bf 23}, (1969) 851;
D.~R.~Bosomworth, W.~Hayes, A.~R.~L.~Spray and G.~D.~Watkins, Proc.~R.~Soc.~A, {\bf 317}, (1970) 133.

\bibitem{Kaneta} H.~Yamada-Kaneta, C.~Kaneta and T.~Ogawa, 
Phys.~Rev.~B {\bf 42}, 9650 (1990).

\bibitem{ShimuraSi} F.~Shimura, {\it Oxygen in Silicon},
Semiconductors and Semimetals, Vol.~42 (Academic Press, New York, 1994)
and references therein.

\bibitem{Kaneta2}
J.~Kato, K.~M.~Itoh, H.~Yamada-Kaneta and H-J.~Pohl,
Phys.~Rev.~B {\bf 68}, 035205 (2003).

\bibitem{Kaiser} W.~Kaiser, J.~Phys.~Chem.~Solids {\bf 23}, 255 (1962).

\bibitem{Corbett} J.~W.~Corbet, R.~S.~McDonald, and G.~D.~Watkins,
J.~Phys.~Chem.~Solids {\bf 25}, 873 (1964).

\bibitem{Clauws} P.~Clauws, Mater.~Sci.~Eng., B {\bf 36}, 213 (1996).

\bibitem{Pesola}
M.~Pesola, J.~von Boehm, V.~Sammalkorpi, T.~Mattila, and R.~M.~Nieminen,
Phys.~Rev.~B {\bf 60}, R16267 (1999).

\bibitem{Alt} H.~Ch.~Alt, Appl.~Phys.~Lett. {\bf 54}, 1445 (1989);
{\bf 55}, 2736 (1989); Phys.~Rev.~Lett. {\bf 65}, 3421 (1990).

\bibitem{Schneider}
J.~Schneider, B.~Dischler, H.~Seelewind, P.~M.~Mooney, J.~Lagowski, M.~Matsui,
D.~R.~Beard, and R.~C.~Newman, Appl.~Phys.~Lett. {\bf 54}, 1442 (1989).


\bibitem{Gienger} M.~Gienger, M.~Glaser and K.~Lassmann, 
Solid State Commun. {\bf 86}, 285 (1993).


\bibitem{Mayur} A.~J.~Mayur, M.~D.~Sciacca, M.~K.~Udo, 
A.~K.~Ramdas, K.~Itoh, J.~Wolk, and E.~E.~Haller,
Phys.~Rev.~B {\bf 49}, 16293 (1994).


\bibitem{Artacho97} 
E.~Artacho, F.~Yndur\'ain, B.~Pajot, R.~Ram\'irez, C.~P.~Herrero, L.~I.~Khirunenko,
K.~M.~Itoh, and E.~E.~Haller,
Phys. Rev. B {\bf 56}, 3820 (1997).


\bibitem{Pajot2000} B.~Pajot, P.~Clauws, J.~L.~Lindstr\"om, and E.~Artacho,
Phys.~Rev.~B {\bf 62}, 10165 (2000).


\bibitem{Coutinho} 
J.~Coutinho, R.~Jones, P.~R.~Briddon and S.~\"Oberg,
Phys. Rev. B {\bf 62}, 10824 (2000).


\bibitem{PRB} H.~Shima and T.~Nakayama, Phys.~Rev.~B {\bf 69}, 035202 (2004).

\bibitem{PRA} H.~Shima and T.~Nakayama, Phys.~Rev.~A {\bf 70}, 013401 (2004).

\bibitem{Esqui} See, {\it Tunneling Systems in Amrphous and Crystalline Solids},
edited by P.~Esquinazi (Springer-Verlag, Berlin 1998).


\bibitem{foot} 
An exact value of $|\bm{p}|$ may be obtained 
through a quantum-chemical calculation;
to perform this is beyond the present study.
The influence of the change in the definition of $|\bm{p}|$
to the results reached in this work
will be presented at need.

\bibitem{Whittaker} E.~T.~Whittaker and G.~N.~Watson,
{\it A course of modern analysis}, (Cambridge University Press 1962).


\bibitem{Gloden} R.~F.~Gloden, Euratom Rpt. EUR 4349f (1970).


\bibitem{Wurger} See, A.~W\"urger, {\it Coherent Tunneling to Relaxation},
(Springer-Verlag, 1997).


\bibitem{Solf_Klein}
M.~P.~Solf and M.~W.~Klein, Phys.~Rev.~B {\bf 46}, 8147 (1992); {\bf 47}, 11097 (1993).


\bibitem{AHV} P.~W.~Anderson, B.~I.~Halperin, and C.~M.~Varma, Phil. Mag. {\bf 25}, 1 (1972).


\bibitem{Phillips} W.~A.~Phillips, J.~Low.~Temp.~Phys. {\bf 7}, 351 (1972).

\bibitem{Landau} L.~D.~Landau and E.~M.~Lifshitz,
{\it Quantum Mechanics}, (Pergamon Press Oxford, 1977).

\bibitem{Schnelle} W.~Schnelle and E.~Gmelin,
J.~Phys.~Condens.~Matter {\bf 13}, 6087 (2001).


\bibitem{Mehta} See, M.~L.~Mehta, {\it Random Matrices} (Academic Press, NY, 1991),
for derivating the explicit form of $P(R)$ in one-dimensional systems.


\bibitem{Klein} M.~W.~Klein, Phys.~Rev.~B {\bf 29}, 5825 (1984);
{\bf 31}, 1114 (1985).


\bibitem{Enss} C.~Enss, M.~Gaukler, S.~Hunklinger, M.~Tornow, 
R.~Weis, and A. W\"urger,
Phys.~Rev.~B {\bf 53}, 12094 (1996).


\bibitem{Wurger_Euro} A.~W\"urger, R.~Weis, M.~Gaukler and C.~Enss,
Europhys.~Lett. {\bf 33}, 533 (1996).


\bibitem{Wurger1}
A.~W\"urger, Z.~Phys.~B {\bf 94}, 173 (1994); {\bf 98}, 561 (1995).

\bibitem{Wurger2}
O.~Terzidis and A.~W\"urger, Z.~Phys.~B {\bf 94}, 341 (1994).


\bibitem{foot_chi}
In our previous paper \cite{JPSJ},
the contribution of paired rotors to $\chi(T)$ was overestimated;
this is due to that all rotors contained in Ge:O were factorized
into pairs regardless of a random distribution of rotors in a cubic space.

\bibitem{JPSJ} H.~Shima and T.~Nakayama, J.~Phys.~Soc.~Jpn {\bf 73}, 2464 (2004).


\bibitem{Bara} S.~D.~Baranovskii, B.~I.~Shklovskii, and A.~L.~Efros, 
Sov.~Phys. JETP {\bf 51}, 199 (1980).


\bibitem{Efros} A.~L.~Efros and B.~I.~Shklovskii,
J.~Phys.~C {\bf 8}, L49 (1975);
in {\it Electron-Electron Interaction in Disordered Systems},
edited by A.~L.~Efros and M.~Pollak,
Elsevier Science Publications B.V. (1985).


\bibitem{Kirk} S.~K.~Kirkpatrick and C.~M.~Varma,
Solid State Commun. {\bf 25}, 821 (1978).


\bibitem{Fisch} R.~Fisch, Phys.~Rev.~B {\bf 22}, 3459 (1980).



\bibitem{Zeller} R.~C.~Zeller and R.~O.~Pohl, Phys. Rev. B {\bf 4}, 2029 (1971).



\bibitem{Salvino} D.~J.~Salvino, S.~Rogge, B.~Tigner and D.~D.~Osheroff,
Phys. Rev. Lett. {\bf 73}, 268 (1994).


\bibitem{Burin} A.~L.~Burin, J.~Low.~Temp.~Phys. {\bf 100}, 309 (1995);
A.~L.~Burin and Y.~Kagan, JETP {\bf 80}, 761 (1995).


\bibitem{Osheroff} 
D.~Rosenberg, P.~Nalbach and D.~D.~Osheroff,
Phys.~Rev.~Lett. {\bf 90}, 195501 (2003);
S.~Ludwig, P.~Nalbach, D.~Rosenberg, and D.~Osheroff,
Phys. Rev. Lett. {\bf 90}, 105501 (2003);
S.~Ludwig and D.~D.~Osheroff, Phys.~Rev.~Lett. {\bf 91}, 105501 (2003).


\bibitem{Ladieu} J.~Le~Cochec and F.~Ladieu, Eur.~Phys.~J.~B {\bf 32}, 13 (2003);
F.~Ladieu, J.~Le~Cochec, P.~Pari, P.~Trouslard, and P.~Ailloud 
Phys.~Rev.~Lett. {\bf 90}, 205501 (2003).


\end{thebibliography}
\end{document}